\documentclass[journal]{IEEEtran}
\usepackage[utf8]{inputenc}
\usepackage{bm}
\usepackage{amsfonts,amsmath,amssymb}
\usepackage{graphicx} 
\usepackage{soul}
\usepackage{mwe}
\usepackage[binary-units=true]{siunitx}

\usepackage{booktabs}
\usepackage{multirow}
\usepackage[table,xcdraw]{xcolor}

\usepackage{pgfplots}
\usetikzlibrary{decorations, arrows}
\pgfplotsset{compat=1.17}

\usepackage[acronym,shortcuts]{glossaries}
\newacronym{arq}{ARQ}{automatic repeat request}
\newacronym{cdf}{CDF}{cumulative distribution function}
\newacronym{cir}{CIR}{channel impulse response}
\newacronym{ecg}{ECG}{electrocardiogram}
\newacronym{lrt}{LRT}{likelihood ratio test}
\newacronym{mi}{MI}{mutual information}
\newacronym{ml}{ML}{machine learning}
\newacronym{mse}{MSE}{mean square error}
\newacronym{nn}{NN}{neural network}
\newacronym{kde}{KDE}{kernel density estimation}
\newacronym{pls}{PLS}{physical-layer security}
\newacronym{pdf}{PDF}{probability density function}
\newacronym{rms}{RMS}{root mean square}
\newacronym{rng}{RNG}{random number generator}
\newacronym{ska}{SKA}{secret-key-agreement}
\newacronym{skr}{SKR}{secret key rate}
\newacronym{uwac}{UWAC}{underwater acoustic channel}
\newacronym{uwan}{UWAN}{underwater acoustic network}
\newacronym{wban}{WBAN}{wireless body area network}
\newacronym{adqc}{ADQC}{advantage distillation with qunantization correction}

\DeclareMathOperator*{\argmax}{arg\,max}
\DeclareMathOperator*{\argmin}{arg\,min}

\usepackage{pifont}

 \newlength\fwidth
\newlength\fheight
\usepackage[caption=false]{subfig}

\title{Secret-Key-Agreement Advantage Distillation \\ With   Quantization Correction}
\author{Francesco Ardizzon, {\em IEEE Member}, Francesco Giurisato, and Stefano Tomasin, {\em IEEE Senior Member}\thanks{Manuscript received --; accepted --. Date --; date of current version --. Corresponding author: F. Ardizzon. The authors are with the Department of Information Engineering, Universit\`a degli Studi di Padova, Padua 35131, Italy. S. Tomasin is also with the National Inter-University Consortium for Telecommunications (CNIT), 43124 Parma, Italy. (email:  francesco.ardizzon@phd.unipd.it, francesco.giurisato@studenti.unipd.it, stefano.tomasin@unipd.it). }}

\begin{document}

\maketitle
\begin{abstract}
    We propose a novel advantage distillation strategy for physical layer-based \ac{ska}. We consider a scenario where Alice and Bob aim at extracting a common bit sequence, which should remain secret to Eve, by quantizing a random number obtained from measurements at their communication channel. We propose an asymmetric advantage distillation protocol with two novel features: i) Alice quantizes her measurement and sends partial information on it over an authenticated public side channel, and ii) Bob quantizes his measurement by exploiting the partial information. The partial information on the position of the measurement in the quantization interval and its sharing allows Bob to obtain a quantized value closer to that of Alice. Both strategies increase the lower bound of the secret key rate.
    %Numerical results prove the effectiveness of our approach, showing that exchanging few bits of information before the actual information reconciliation allows to extract more secret bits from the channel.
\end{abstract}


\glsresetall
\sloppy

\section{Introduction}
\Ac{ska} is a security mechanism by which two users, namely Alice and Bob, agree on a common key while keeping it secret from any third malicious user, namely Eve. The secret key can then be used for other security services, e.g., for symmetric key encryption or authentication.

Initially proposed by Maurer \cite{maurer93}, Ahlswede, and Csiszar \cite{ahlswede93}, physical-layer-based \ac{ska} schemes are information-theoretic secure, and their security is based on the physical properties of the channel itself. A source-model \ac{ska} procedure involves four steps \cite{bloch}: {\em channel probing}, where Alice and Bob transmit in turn probing signals and collect the channel measurements later used to extract the keys; {\em advantage distillation} by which each agent extracts a bit sequence from his/her measurement; {\em information reconciliation}, where Alice and Bob exchange information with the aim of reducing the disagreement among the bit sequences; finally, {\em privacy amplification}, where each user extracts from the bit sequences a shorter one typically by using universal hashing (for further details see surveys \cite{Jorswieck15} and \cite{li19}).

% STATE-OF-THE-ART- ADV. DISTILLATION
In this paper, we focus on the advantage distillation step. The basic approach requires quantizing the channel feature used for key extraction.
A channel quantization scheme for multiple-input multiple-output (MIMO) channels is proposed in \cite{chen10} and \cite{Chen2011}. In particular, in the strategy of \cite{chen10} Alice transmits a quantization correction to Bob, the observations have a (known) Gaussian distribution, and the quantizer thresholds are set to provide equiprobable bit sequences (with maximum entropy). However, Eve's observations are assumed to be independent of those of Bob. We consider here instead a more realistic scenario, where the features' distribution is not known a priori, and Eve's observations are statistically correlated to those of Alice and Bob.

 In \cite{graur16} the quantization intervals are separated by guard bands and samples falling in these regions are discarded to reduce quantization mismatches between Alice and Bob. Indeed, this increases the probability of agreement and the bit sequence length, at the expense of fewer extracted bits. A related approach is also proposed in \cite{Adil21}, where the quantizer thresholds are set to assure that each sequence is equiprobable, maximizing the output entropy. In both works, legitimates' and Eve's channels are assumed to be uncorrelated, thus no information about the actual bit sequence by collected from Eve.

Recently, a technique to extract bits from \acp{ecg} signals for \acp{wban} has been proposed in \cite{Guglielmi21}. The quantizer thresholds are optimized to maximize both the entropy and the matching rate of the extracted bits. Still, due to the particular nature of the channel, no information is leaked to Eve during the channel probing step. We consider instead the case wherein Eve is observing a channel correlated to that of Alice and Bob, and Eve also overhears any public discussion between Alice and Bob.

%Considering the channel model-\ac{ska}, an advantage distillation that exploits information exchange between Alice and Bob is proposed in \cite{tomasin14}. In particular, the authors proposed to discard bits that are not reliable at Bob, i.e., with a low log-likelihood ratio. \hl{FA: \'E channel model: teniamo o togliamo?}  %In \cite{park20}, a similar approach is proposed, based instead on code scrambling.

In this letter, we propose a novel advantage distillation strategy for a source-model \ac{ska}, where Alice and Bob obtain each a random number and optimize their quantizers to obtain bit sequences providing the highest secret key rate (through a lower bound). Then, they coordinate the quantization of the observed feature with a discussion over a public authenticated channel. In particular, Alice quantizes her measurement and sends the position of the measurement in the quantization interval over an authenticated public side channel. In turn, Bob (and Eve) quantizes his measurement by exploiting the partial information. We denote the described advantage distillation technique as \ac{adqc}. We show that such a strategy allows the extraction of more secret bits from the channel measurements. 
Finally, with respect to the existing literature, we show that a careful design of the quantizers used during the advantage distillation and the transmission of quantization error correction over a public channel allows Alice and Bob to obtain a secret key, even in those harsh scenarios where Eve is close to one of the agents.

The rest of the paper is organized as follows. Section \ref{sec:sysModel} introduces the system model. Section \ref{sec:propProtocol} describes the step of the proposed advantage distillation protocol. Section \ref{sec:numResults} presents the numerical results. Section \ref{sec:conclusions} draws the conclusions.

\section{System Model}\label{sec:sysModel}

\begin{figure}
    \centering
    \resizebox{\columnwidth}{!}{\tikzset{every picture/.style={line width=0.75pt}} %set default line width to 0.75pt        

\begin{tikzpicture}[x=0.75pt,y=0.75pt,yscale=-1,xscale=1]

%Shape: Can [id:dp23869138945480795] 
\draw  [color={rgb, 255:red, 74; green, 144; blue, 226 }  ,draw opacity=1 ] (137.62,129.54) -- (91.72,129.3) .. controls (90.11,129.29) and (88.79,124.93) .. (88.77,119.57) .. controls (88.75,114.2) and (90.04,109.86) .. (91.65,109.86) -- (137.55,110.1) .. controls (139.16,110.11) and (140.48,114.47) .. (140.5,119.84) .. controls (140.52,125.2) and (139.23,129.55) .. (137.62,129.54) .. controls (136.01,129.53) and (134.69,125.17) .. (134.67,119.81) .. controls (134.65,114.44) and (135.94,110.1) .. (137.55,110.1) ;
%Straight Lines [id:da16118526220660545] 
\draw    (60.25,90.05) -- (86.44,61.13) ;
\draw [shift={(87.78,59.65)}, rotate = 132.16] [color={rgb, 255:red, 0; green, 0; blue, 0 }  ][line width=0.75]    (4.37,-1.32) .. controls (2.78,-0.56) and (1.32,-0.12) .. (0,0) .. controls (1.32,0.12) and (2.78,0.56) .. (4.37,1.32)   ;
%Shape: Can [id:dp8299862878922792] 
\draw  [color={rgb, 255:red, 208; green, 2; blue, 27 }  ,draw opacity=1 ] (136.59,69.37) -- (90.69,69.37) .. controls (89.08,69.37) and (87.78,65.01) .. (87.78,59.65) .. controls (87.78,54.28) and (89.08,49.93) .. (90.69,49.93) -- (136.59,49.93) .. controls (138.2,49.93) and (139.51,54.28) .. (139.51,59.65) .. controls (139.51,65.01) and (138.2,69.37) .. (136.59,69.37) .. controls (134.98,69.37) and (133.68,65.01) .. (133.68,59.65) .. controls (133.68,54.28) and (134.98,49.93) .. (136.59,49.93) ;
%Straight Lines [id:da6653588316833712] 
\draw    (136.5,60.55) -- (163.25,60.5) ;
\draw [shift={(165.25,60.5)}, rotate = 179.9] [color={rgb, 255:red, 0; green, 0; blue, 0 }  ][line width=0.75]    (4.37,-1.32) .. controls (2.78,-0.56) and (1.32,-0.12) .. (0,0) .. controls (1.32,0.12) and (2.78,0.56) .. (4.37,1.32)   ;
%Straight Lines [id:da3111003136218986] 
\draw    (60.25,90.05) -- (87.38,118.13) ;
\draw [shift={(88.77,119.57)}, rotate = 225.98] [color={rgb, 255:red, 0; green, 0; blue, 0 }  ][line width=0.75]    (4.37,-1.32) .. controls (2.78,-0.56) and (1.32,-0.12) .. (0,0) .. controls (1.32,0.12) and (2.78,0.56) .. (4.37,1.32)   ;
%Straight Lines [id:da8620638306622488] 
\draw    (137.5,119.8) -- (164.25,119.75) ;
\draw [shift={(166.25,119.75)}, rotate = 179.9] [color={rgb, 255:red, 0; green, 0; blue, 0 }  ][line width=0.75]    (4.37,-1.32) .. controls (2.78,-0.56) and (1.32,-0.12) .. (0,0) .. controls (1.32,0.12) and (2.78,0.56) .. (4.37,1.32)   ;
%Shape: Can [id:dp716874305660292] 
\draw  [color={rgb, 255:red, 74; green, 144; blue, 226 }  ,draw opacity=1 ] (275.28,109.72) -- (321.18,109.72) .. controls (322.79,109.72) and (324.1,114.07) .. (324.1,119.43) .. controls (324.1,124.8) and (322.79,129.15) .. (321.18,129.15) -- (275.28,129.15) .. controls (273.67,129.15) and (272.37,124.8) .. (272.37,119.43) .. controls (272.37,114.07) and (273.67,109.72) .. (275.28,109.72) .. controls (276.89,109.72) and (278.2,114.07) .. (278.2,119.43) .. controls (278.2,124.8) and (276.89,129.15) .. (275.28,129.15) ;
%Straight Lines [id:da29006780971821433] 
\draw    (351.6,89.28) -- (325.24,61.58) ;
\draw [shift={(323.86,60.13)}, rotate = 46.42] [color={rgb, 255:red, 0; green, 0; blue, 0 }  ][line width=0.75]    (4.37,-1.32) .. controls (2.78,-0.56) and (1.32,-0.12) .. (0,0) .. controls (1.32,0.12) and (2.78,0.56) .. (4.37,1.32)   ;
%Shape: Can [id:dp10677980192375269] 
\draw  [color={rgb, 255:red, 208; green, 2; blue, 27 }  ,draw opacity=1 ] (275.04,50.41) -- (320.94,50.41) .. controls (322.55,50.41) and (323.86,54.77) .. (323.86,60.13) .. controls (323.86,65.5) and (322.55,69.85) .. (320.94,69.85) -- (275.04,69.85) .. controls (273.43,69.85) and (272.13,65.5) .. (272.13,60.13) .. controls (272.13,54.77) and (273.43,50.41) .. (275.04,50.41) .. controls (276.65,50.41) and (277.96,54.77) .. (277.96,60.13) .. controls (277.96,65.5) and (276.65,69.85) .. (275.04,69.85) ;
%Straight Lines [id:da7157308935093365] 
\draw    (351.6,89.28) -- (325.45,117.96) ;
\draw [shift={(324.1,119.43)}, rotate = 312.37] [color={rgb, 255:red, 0; green, 0; blue, 0 }  ][line width=0.75]    (4.37,-1.32) .. controls (2.78,-0.56) and (1.32,-0.12) .. (0,0) .. controls (1.32,0.12) and (2.78,0.56) .. (4.37,1.32)   ;
%Straight Lines [id:da6397777868694732] 
\draw    (248.75,59.78) -- (275.5,59.73) ;
\draw [shift={(246.75,59.78)}, rotate = 359.9] [color={rgb, 255:red, 0; green, 0; blue, 0 }  ][line width=0.75]    (4.37,-1.32) .. controls (2.78,-0.56) and (1.32,-0.12) .. (0,0) .. controls (1.32,0.12) and (2.78,0.56) .. (4.37,1.32)   ;
%Straight Lines [id:da31072012869305743] 
\draw    (248.5,120.03) -- (275.25,119.98) ;
\draw [shift={(246.5,120.03)}, rotate = 359.9] [color={rgb, 255:red, 0; green, 0; blue, 0 }  ][line width=0.75]    (4.37,-1.32) .. controls (2.78,-0.56) and (1.32,-0.12) .. (0,0) .. controls (1.32,0.12) and (2.78,0.56) .. (4.37,1.32)   ;
%Straight Lines [id:da7696729669087072] 
\draw [color={rgb, 255:red, 208; green, 2; blue, 27 }  ,draw opacity=1 ] [dash pattern={on 4.5pt off 4.5pt}]  (200.25,10.25) -- (200.25,150.5) ;

% Text Node
\draw (167.5,53) node [anchor=north west][inner sep=0.75pt]   [align=left] {Bob};
% Text Node
\draw (100,33) node [anchor=north west][inner sep=0.75pt]   [align=left] {${h}_{\rm AB}$};
% Text Node
\draw (19.65,81) node [anchor=north west][inner sep=0.75pt]   [align=left] {Alice};

\draw (167.5,112) node [anchor=north west][inner sep=0.75pt]   [align=left] {Eve};
\draw (100,92) node [anchor=north west][inner sep=0.75pt]   [align=left] {${h}_{\rm AE}$};
\draw (354.15,81) node [anchor=north west][inner sep=0.75pt]   [align=left] {Bob};
\draw (285,33) node [anchor=north west][inner sep=0.75pt]   [align=left] {${h}_{\rm BA}$};
\draw (210.5,53) node [anchor=north west][inner sep=0.75pt]   [align=left] {Alice};

\draw (217.15,112) node [anchor=north west][inner sep=0.75pt]   [align=left] {Eve};
\draw (285,92) node [anchor=north west][inner sep=0.75pt]   [align=left] {${h}_{\rm BE}$};

\end{tikzpicture}}
    \caption{Scheme of a channel probing procedure.}
    \vspace{-.6cm}
    \label{fig:chProbing}
\end{figure}
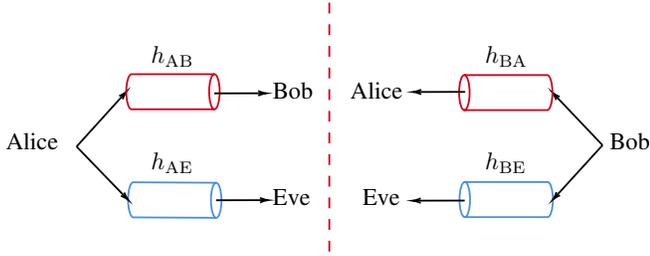

We consider a scenario where Alice and Bob aim to agree on a common bit sequence, which has to stay secret from Eve. To this end, they use a source model \ac{ska} procedure \cite{bloch}. First, they probe their channel, as shown in  Fig.~\ref{fig:chProbing}: Alice and Bob alternatively send pilot signals through the connecting wireless channel to enable their partner to estimate the channel, so that Alice obtains the estimated channel $h_\mathrm{BA}$ and Bob obtains estimated channel $h_\mathrm{AB}$.

We assume Alice and Bob have already agreed on a feature selection and extraction function such that Alice extracts $x$ from $h_\mathrm{BA}$, while Bob extracts $y$ from $h_\mathrm{AB}$. We focus on the scalar case where $x$ and $y$ are real numbers, although the SKA will operate on sequences of $x$ and $y$, thus using longer observation sequences. We remark that, in general, the channels are only partially reciprocal, therefore $x$ and $y$ will be strongly correlated but not identical. %\hl{se avessimo vettori, potremmo sfruttare ad es sottospazi ignoti a Eve, molto piu' complicato, si passerebbe a vector quantization}

Eve is modeled as a passive attacker. From each exchange, she estimates channels $h_\mathrm{AE}$ and $h_\mathrm{BE}$, from Alice and Bob, respectively.  We assume Eve has an extraction function that exploits (one or) both channels and retrieves the scalar real feature ${z}$. Indeed, if Eve and Bob (or Alice) are in a different position, $y\neq z$ and $x\neq z$. Still, if Eve is not too far from Alice or Bob, there exists a non-negligible correlation between $z$ and both $x$ and $y$.

We assume that the statistics of $x$, $y$, and $z$ are not known in close form, but a dataset of measurements is available to all parties for the design of the SKA procedure.

An authenticated public side channel is available, over which Alice and Bob can exchange information, while Eve overhears any communication. Channel coding is used on this side channel, allowing Bob to detect and correct, with arbitrarily small error probability, any error of publicly exchanged information.

\section{Advantage Distillation \\ With Quantization Correction}\label{sec:propProtocol}

We now describe the \ac{adqc} technique. Let us introduce the binary space $\mathcal{S} = \{0,1\}^{b}$  containing $M = 2^{b}$ different binary strings, each of $b$ bits. Alice and Bob aim at drawing two sequences, $\bm{s}_\mathrm{A} \in \mathcal{S}$ and $\bm{s}_\mathrm{B}\in  \mathcal{S}$, by processing the observed channel features $x$ and $y$, respectively.

The problem of associating a real number (in this case, the feature measurement) to a binary sequence can be seen as a quantization problem that partitions the set of real numbers into $M$ intervals so that the $m$-th interval is  associated with the sequence $\bm{s}_m \in \mathcal{S}$. A quantizer $q$ provides the bit sequence $\bm{s}=q(a)$ from the real number $a$.

First, note that the quantizers used by Alice, Bob, and Eve are chosen before the actual key agreement protocol, as will be detailed later. Moreover, we consider a worst-case scenario where all quantizers are publicly known, However,    both the secrecy and the randomness of the scheme still lie in the extracted channel measurements. 

Now, we aim to make this extraction process such that $\bm{s}_\mathrm{A}$ is as close as possible to $\bm{s}_\mathrm{B}$ while remaining secret to Eve. We can write the observation at Bob as the observation at Alice corrupted by an error $\epsilon$, i.e.,
\begin{equation} \label{yvsx}
y = x + \epsilon. 
\end{equation}
Let $q_{\rm A}(x)$ be the quantized value at Alice (corresponding to the $m$-th quantization interval), and let $\eta = x- q(x)$ be the quantization error at Alice. Then, from \eqref{yvsx} we have
\begin{equation}
y = q_{\rm A}(x) + \eta + \epsilon.
\end{equation}
In general, note that $\eta$ and $\epsilon$ are statistically dependent. However, ignoring this dependency, we can have that $y$ is turned away from the quantization value $q_{\rm A}(x)$ by both errors $\eta$ and $\epsilon$. Thus, to improve the advantage distillation procedure, in  \ac{adqc} Alice communicates over the public channel the value of the quantization error $\eta$ so that Bob can compute
\begin{equation}\label{eq:bobCorrection}
y' = y-\eta = q_{\rm A}(x) + \epsilon,
\end{equation}
and quantize $y'$ with quantizer $q_{\rm B}(\cdot)$ to obtain its bit sequence.

If Alice uses $B$ bits to feedback $\eta$ over the public channel, we must quantize $\eta$. To this end, each quantization interval $\mathcal I^{\rm (A)}_m$, $m=1, \ldots, M$,    is split into $K=2^B$ sub-intervals of equal length, and (a binary representation) of the index of the sub-interval in which $\eta$ is falling is transmitted over the public channel.  Then, Alice transmits
\begin{equation}\label{eq:scaling}
    \xi = \left\lceil\eta\frac{K}{L^{\rm (A)}(x)}\,  \right\rceil \;,
\end{equation}
where $L^{\rm (A)}(x)$ is the length of the quantization interval of $x$. This quantization procedure also avoids transmitting the value of $\eta$ that may reveal in part the interval $\mathcal I_m$ to Eve, since quantization intervals may have different lengths.

% Moreover, since in general, the quantization intervals have different sizes, also values of quantization error $\eta$ have different ranges, and transmitting it on the public channel may reveal information on the quantization interval (thus the extracted bit sequence) to Eve. Therefore, instead of transmitting $\eta$ Alice transmits  
% \begin{equation}\label{eq:scaling}
%     \xi = \left\lceil\eta\frac{K}{L_m}\,  \right\rceil \;,
% \end{equation}
% where $L_m$ is the length of the $m$-th quantization interval and we have split the quantization interval into $K = 2^{B}-1$ sub-bands having the same size, each identified by index $k$.

Upon reception of $\xi$, Bob computes
\begin{equation}\label{eq:rescaling}
  {\eta}' =\xi  \, L^{\rm (B)}(y)\;,
\end{equation}
where $L^{\rm (B)}(y)$ is the length of quantization interval of $y$. Then Bob uses ${\eta}'$ instead of $\eta$ in \eqref{eq:bobCorrection} to quantize $y'$ with $q_{\rm B}(\cdot)$. Indeed, it may happen that $L^{\rm (A)}(x) \neq L^{\rm (B)}(y)$. Nonetheless, it is reasonable to assume the length of intervals close to each other to be similar.  

Eve can do the same procedure of Bob, by computing its own correction factor $\eta''$ and applying it to its measurement $z$ before quantizing it with $q_{\rm E}(\cdot)$. However, there will be a higher probability that $z' = z - \eta''$ falls in another interval than $x$, thus the correction factor won't provide the same benefit on the sequence extraction of Bob.

%are correlated, they will end up to intervals associated with the same bit sequence and thus, reasonably, with the size or to intervals that are close the each other. Still, in this latter case, it is reasonable to assume that, for high values of $b$ intervals close to each other will have similar sizes, thus still allowing the recovery of the correct bit sequence.

%We propose an advantage distillation protocol where, Alice and Bob agree on their quantizers, $q_\mathrm{A}$ and $q_\mathrm{B}$, via public discussion. In particular, for each collected pair of channel features, $(x,y)$,  
%\begin{enumerate}
%    \item Alice quantizes her own measurement to obtain the bit sequence 
%    \begin{equation}
%        \bm{s}_\mathrm{A} = q_\mathrm{A}(x) \;.
%    \end{equation}
%    \item Alice computes a feature correction, $\xi$, and transmits it via a public (authenticated) channel to Bob;
%    \item Bob exploits the correction $\xi$ to to obtain the bit sequence as
%    \begin{equation}\label{eq:bobCorrection}
%        \bm{s}_\mathrm{B} = q_\mathrm{B}(y') = q _\mathrm{B}(y - \xi )\;.
%    \end{equation}
%\end{enumerate}
%Indeed, at the end of the process, we aim $\bm{s}_\mathrm{B}$ to be as similar as possible to $\bm{s}_\mathrm{A}$.

%Eve can exploit the correction $\xi$ and her own measurement $z$, extracting the sequence
%\begin{equation}
%    z' = z - \xi\;,
%\end{equation}
%\begin{equation}
%    \bm{s}_\mathrm{E} = q_\mathrm{E}( z' )  =  q_\mathrm{E}( z - \xi) \;.
%\end{equation}

\subsection{Quantizer Design}
We are now left with the design of the Alice, Bob, and Eve quantizers, i.e., $q_\mathrm{A}$, $q_\mathrm{B}$, and  $q_\mathrm{E}$, respectively.

Now, note that a quantizer $q$ with $M$ quantization intervals is fully defined by the position of $M+1$ thresholds, $\mathcal{T} = \{T_i, i=0, \ldots, M+1\}$, where however the saturation values $T_0 = T_\mathrm{min}$ and $T_{M+1}=T_\mathrm{max}$ are set to match a predefined saturation probability. \footnote{Samples eventually falling outside the region $[T_\mathrm{min},T_\mathrm{max}]$ are remapped to the closest interval.} Let $\mathcal{T}_\mathrm{A}$, $\mathcal{T}_\mathrm{B}$, and $\mathcal{T}_\mathrm{E}$ be sets of thresholds used for the three quantizers. The metric used for the design is the lower bound on the secret-key capacity for the source model  \cite[Ch.\,4]{maurer93,bloch}, i.e.,
\begin{equation}\begin{split}\label{eq:obj_general}
        C_\mathrm{sk}^\mathrm{low}(\mathcal{T}_\mathrm{A},\mathcal{T}_\mathrm{B},\mathcal{T}_\mathrm{E}) = 
        I(\bm{s}_\mathrm{A};\bm{s}_\mathrm{B}) - \min \left\{ I(\bm{s}_\mathrm{A}; \bm{s}_\mathrm{E}), I(\bm{s}_\mathrm{B}; \bm{s}_\mathrm{E}) \right\}.
        %C_\mathrm{sk}^\mathrm{low} &=  \max \left(  I(\bm{s}_\mathrm{A};\bm{s}_\mathrm{B}) - I(\bm{s}_\mathrm{A}; \bm{s}_\mathrm{E}) ,  I(\bm{s}_\mathrm{A};\bm{s}_\mathrm{B}) - I(\bm{s}_\mathrm{B}; \bm{s}_\mathrm{E}) \right) = \\
        %= I(q_\mathrm{A}(x);q_\mathrm{B}(y')) - \min \left\{ I(q_\mathrm{A}(x);q_\mathrm{E}(z')), I(q_\mathrm{A}(x);q_\mathrm{E}(z')) \right\},
\end{split}    
\end{equation}
where $I(\bm{v}_1;\bm{v}_2)$ is the mutual information between random vectors $\bm{v}_1$ and $\bm{v}_2$. Alice and Bob aim at designing the quantizers $q_\mathrm{A}$ and $q_\mathrm{B}$ to increase $C_\mathrm{sk}^\mathrm{low}$, i.e., by increasing the agreement between Alice's and Bob's extracted bit sequences, while limiting the amount of information revealed to Eve. Eve in turn aims at minimizing $C_\mathrm{sk}^\mathrm{low}(\mathcal{T}_\mathrm{A},\mathcal{T}_\mathrm{B},\mathcal{T}_\mathrm{E})$ with a proper choice of her quantizer $q_\mathrm{E}$.

To estimate the mutual information it is necessary to have the associated joint \ac{pdf}: this is either known a priori or estimated by using a dataset of observations $(x,y,z)$ as input to the quantizers.

%For instance, the quantized values associated by Alice with the $m$-th interval is  
%\begin{equation}
%    c_m \triangleq \int_{T_m}^{T_{m+1}} a\, p_x(a) da  \;,
%\end{equation}
%where $p_x(a)$ is the \ac{pdf} of Alice measurement. 

%Next, the length of interval $\mathcal{I}_m$ turns out to be 
%\begin{equation}\label{eq:intervalsLen}
%     L_m \triangleq 
%    \begin{cases}
%    2\left(T_1 - c_1\right) & \mbox{if } m= 1, \\
%    T_{m+1} - T_{m} & \mbox{if } 1<m<M,\\
%    2\left(c_M - T_{M}\right) & \mbox{if }  m=M. \\
%    \end{cases}
%\end{equation}

To design the quantizer we consider the following iterative procedure. Starting from uniform quantizers on a predefined range, at each iteration Eve optimizes her quantizer
\begin{equation}\label{eq:Eve_optim}
        \hat{\mathcal{T}}_\mathrm{E} = \argmin_{\mathcal{T}_\mathrm{E}} C_\mathrm{sk}^\mathrm{low}(\mathcal{T}_\mathrm{A},\mathcal{T}_\mathrm{B},\mathcal{T}_\mathrm{E})\;,
\end{equation}
with $\mathcal{T}_\mathrm{A}$ and $\mathcal{T}_\mathrm{B}$ fixed.
Next, Alice and Bob optimize their own
\begin{equation}\label{eq:AliceBob_optim}
    [\hat{\mathcal{T}_\mathrm{A}},\hat{\mathcal{T}_\mathrm{B}}] = \argmax_{{\mathcal{T}_\mathrm{A}},{\mathcal{T}_\mathrm{B}}} C_\mathrm{sk}^\mathrm{low}(\mathcal{T}_\mathrm{A},\mathcal{T}_\mathrm{B},\hat{\mathcal{T}}_\mathrm{E})\;.
\end{equation}
Finally, Alice, Bob, and Eve set the quantizers $\hat{q}_\mathrm{A}$, $\hat{q}_\mathrm{B}$, and $\hat{q}_\mathrm{E}$, from the new thresholds $\hat{\mathcal{T}}_\mathrm{A}$, $\hat{\mathcal{T}}_\mathrm{B}$, and $\hat{\mathcal{T}}_\mathrm{E}$.  The optimizations are performed via numerical methods. The procedure is repeated either until convergence is reached or a maximum number of iterations has been performed.

\subsection{Advantage Distillation vs Information Reconciliation with Limited-Rate Public Channel}\label{sec:reconcile}

When the public channel has no rate limitations, a large value of $B$ (number of bits describing the quantization error) is to be preferred to improve the agreement between the bit sequences extracted by Alice and Bob. However, in a scenario where the side-channel rate is limited, and it is used for both advantage distillation and information reconciliation, we must decide the number of bits to be used for both processes.

For \ac{adqc}, we have seen that $B$ bits are transmitted for each quantized sample.
For the information reconciliation, a sequence of $n>b$ bits obtained from the advantage distillation is considered an error-corrupted version of a codeword of a linear code $(k,n)$ as done, for instance, in~\cite{biham06}. Hence, during the reconciliation, Bob will share $n-k$ bits over the public channel for $n/b$ samples. The number of bits shared on the public channel for each bit of the extracted bit sequence is $ \beta \triangleq \frac{B}{b}$, with $\beta = 0$ when no information is shared during advantage distillation, in what we will denote as no error correction (NEC) technique.

Next, we observe that the code rate is related to the secret key capacity (after the advantage distillation) as follows
\begin{equation}
    C_\mathrm{AB} =\frac{ I(\bm{s}_\mathrm{A}; \bm{s}_\mathrm{B})}{b} = \frac{k}{n}.
\end{equation}
%Next, considering an ideal privacy amplification procedure, the number of secret bits extracted from the sequence of $k$ bits is 
%\begin{equation}
%   \frac{C_\mathrm{sk}^\mathrm{low}}{C_\mathrm{AB}} k = C_\mathrm{sk} n \;.
%\end{equation}

We introduce now the cost function $\gamma$ representing the ratio between the numbers of bits shared on the side channel for the \ac{adqc} and the NEC techniques. For the same number of measurements (thus for the same $n$), the ADQC and NEC techniques generate $k^{\rm (ADQC)}$ and $k^{\rm (NEC)}$ bits of the secret key, respectively. Then,  $\gamma$ is  computed as 
\begin{equation}
    \gamma \triangleq \frac{n-k^{\rm  (ADQC)} + \beta n}{n-k^{\rm (NEC)}} = \frac{ 1+\beta  - C_\mathrm{AB}^{\rm (ADQC)}}{ 1  - C_\mathrm{AB}^{\rm (NEC)}},
\end{equation}
where $C_\mathrm{AB}^{\rm (ADQC)}$ and $C_\mathrm{AB}^{\rm (NEC)}$ are  the mutual information between Alice and Bob bit sequences for the ADQC and NEC techniques, respectively.
%Indeed increasing the quantization error correction, guarantees higher mutual information $I(\bm{s}_\mathrm{A}; \bm{s}_\mathrm{B})$, and so an increase of  $C_\mathrm{sk}^\mathrm{low}$, but also increase the cost function $\gamma$. Nonetheless, in a scenario with limited side channel capacity, it is possible to exploit \eqref{eq:sideChannelCost} to find a suitable trade-off between side channel capacity and $C_\mathrm{sk}^\mathrm{low}$.}

%We are now ready to define the merit function $\gamma$, representing the number of obtained secret information bits for each bit transmitted via the authenticated side channel, i.e.,
%{\begin{equation}\label{eq:sideChannelCost}\begin{split}    
 %   \gamma \triangleq & \frac{C_\mathrm{sk}^\mathrm{low} n}{n-k + \beta n }= \frac{C_\mathrm{sk}}{1+\beta -I(\bm{s}_\mathrm{A}; \bm{s}_\mathrm{B}) /b }\;,
%\end{split}
%\end{equation}
% Note that when no information is shared during advantage distillation yields $\beta = 0$.

\section{Numerical Results}\label{sec:numResults}

In this Section, we report the performance of the \ac{adqc} technique and compare it with both the  NEC technique and the  guard-band (GB) technique of \cite{graur16}.

We model the vector $\bm{v} = [x \, y \, z ]^\mathrm{T} $ of Alice's, Bob's, and Eve's measurements as a jointly Gaussian vector having zero-mean and covariance
\begin{equation}
    \bm{\Sigma} = \mathbb{E}[\bm{v} \bm{v}^\mathrm{T}] = 
    \begin{bmatrix}
    1 & \rho_\mathrm{AB} & 0.8 \\
    \rho_\mathrm{AB} & 1 & 0.8 \\
     0.8 & 0.8 &1 \\
    \end{bmatrix}\;,
\end{equation}
where we fixed the correlation between legitimates and Eve features to $\rho_\mathrm{AE} =\rho_\mathrm{BE} = 0.8 $.
Next, we let $\rho_\mathrm{AB}$ varying in the interval  $\rho_\mathrm{AB} \in [0.8, 1]$.
The saturation thresholds are set at $T_\mathrm{max} = -T_\mathrm{min} = 6$, assuring a saturation probability $P_\mathrm{sat} \leq 2\cdot10^{-9}$.

For the \ac{adqc} technique we considered $B = 1$ and $2\,$bit of quantization error correction. For both \ac{adqc} and NEC techniques, quantizers are either optimized as described in the previous section or uniform, with $M-1$ thresholds, placed uniformly in  $[-T_\mathrm{min}, T_\mathrm{max} ]$. For the GB technique, the quantizer is uniform and guard bands are set to $0.85$, to maximize the secret key capacity lower bound.

%\begin{table}[t]
%\centering
%\caption{Summary of the considered SKA techniques}
%\label{tab:scenarios}
%    \begin{tabular}{cccc}\toprule
%                & Quantization Error  & Alice-Bob                & Eve          \\ & Correction &     \\ \midrule
%    A           & \xmark     & Uniform                  & Uniform                  \\
%    \rowcolor[HTML]{EFEFEF} 
%    B           & \xmark     & \cellcolor[HTML]{EFEFEF} & \cellcolor[HTML]{EFEFEF} \\
%    \rowcolor[HTML]{EFEFEF} 
%    C & \cmark & \multirow{-2}{*}{Optimized} & \multirow{-2}{*}{Uniform} \\
%    D           & \xmark     & \multicolumn{1}{c}{}     & \multicolumn{1}{c}{}     \\
%    E & \cmark & \multicolumn{1}{c}{\multirow{-2}{*}{Optimized}}     & \multicolumn{1}{c}{\multirow{-2}{*}{Optimized}}   \\ 
%    \rowcolor[HTML]{EFEFEF}
%    F & \multicolumn{3}{c}{Guard Bands Method \cite{graur16}}\\\bottomrule
%\end{tabular}
%\end{table}

\begin{figure}
    \centering
    \definecolor{mycolor1}{RGB}{77,175,74}%
\definecolor{mycolor2}{RGB}{228,26,28}%
\definecolor{mycolor3}{RGB}{55,126,184}%
\begin{tikzpicture}[every plot/.style={thick}]

\begin{axis}[%
width=.85\columnwidth,
height=.5\columnwidth,
at={(0\fwidth,0\fheight)},
scale only axis = true,
clip marker paths=true,
axis on top=true,
xmin=0.8,
xmax=.999,
xlabel style={font=\color{white!15!black}},
xlabel={$\rho_\mathrm{AB}$},
ymin=0,
ymax=2.3,
xticklabel style={/pgf/number format/fixed, /pgf/number format/precision=3},
xtick distance = 0.02,
ytick distance = 0.5,
ylabel style={font=\color{white!15!black}},
ylabel={$C_\mathrm{sk}^\mathrm{low}\;$[bit]},
axis background/.style={fill=white},
xmajorgrids,
ymajorgrids,
legend style={at={(0.03,0.97)}, anchor=north west, legend cell align=left, align=left, draw=white!15!black},
enlargelimits=false,title style={font=\scriptsize},xlabel style={font=\scriptsize},ylabel style={font=\scriptsize},legend style={font=\scriptsize},ticklabel style={font=\scriptsize}
]

\addplot[]  table[row sep=crcr]{%
0.8	0\\
0.82	0.0538185442724004\\
0.84	0.11324761376503\\
0.86	0.177537772514223\\
0.88	0.250981659024099\\
0.9	0.335289224901301\\
0.92	0.434834313512948\\
0.94	0.55302429998124\\
0.96	0.699184732096944\\
0.97	0.792710621740026\\
0.975	0.851479362602027\\
0.98	0.919965223933813\\
0.985	1.00372911395\\
0.99	1.11163502880957\\
0.995	1.26873583452208\\
0.999	1.53380305130901\\
};
\addlegendentry{NEC unif. quant.}

\addplot [color=mycolor1]
  table[row sep=crcr]{%
0.8	0\\
0.82	0.0618943089462762\\
0.84	0.123249832365641\\
0.86	0.193724106362961\\
0.88	0.27795122946346\\
0.9	0.378169101426133\\
0.92	0.474705203137597\\
0.94	0.625725607944597\\
0.96	0.849740918681548\\
%0.97	0.876493018852354\\
0.975	1.02740413782459\\
%0.98	1.04123076713913\\
0.985	1.22952920738071\\
0.99	1.37347601125521\\
0.995	1.56214973815754\\
0.999	1.90720084322563\\
};
\addlegendentry{NEC, opt. quant.}

\addplot [color=mycolor2, mark = square]
  table[row sep=crcr]{%
0.8	0.0925245681692317\\
0.82	0.154054340571984\\
0.84	0.196189112189851\\
0.86	0.262090598510279\\
0.88	0.272276386707146\\
0.9	0.3935942309924\\
0.92	0.523724488531821\\
0.94	0.698033812214058\\
0.96	0.881869649018066\\
0.97	1.04294800840903\\
0.975	1.14084657567175\\
0.98	1.26497967591259\\
0.985	1.41626328658908\\
0.99	1.61788544235754\\
0.995	1.91635856953453\\
0.999	2.26701823444966\\
};
\addlegendentry{ADQC, $B=\SI{1}{\bit}$}

\addplot [color=mycolor3, mark = o]
  table[row sep=crcr]{%
0.8	0.0855187640195632\\
0.82	0.135474540853017\\
0.84	0.201527201150622\\
0.86	0.264877089608225\\
0.88	0.346509839553604\\
0.9	0.435786427251108\\
0.92	0.555010648990205\\
0.94	0.714085323577574\\
0.96	0.948569640318587\\
0.97	1.11474905160847\\
0.975	1.22542623145895\\
0.98	1.35352246594499\\
0.985	1.5139245893829\\
0.99	1.7309949694044\\
0.995	1.92619429596987\\
0.999	2.28798147287098\\
};
\addlegendentry{ADQC, $B=\SI{2}{\bit}$}

\addplot [color=black, dashed]
  table[row sep=crcr]{%
0.8	0.00759242957590889\\
0.82	0.048072792091719\\
0.84	0.0985104959062983\\
0.86	0.155432985652941\\
0.88	0.215175413268274\\
0.9	0.291588611961264\\
0.92	0.379805472479005\\
0.94	0.493544928232665\\
0.96	0.646141674473278\\
0.97	0.743670203820604\\
0.975	0.807498302226826\\
0.98	0.884616196782038\\
0.985	0.976780039037543\\
0.99	1.11508476064678\\
0.995	1.33640949871437\\
0.999	1.725135498141\\
};
\addlegendentry{GB}

\end{axis}

\end{tikzpicture}%
    \caption{Lower-bound of the secret-key capacity for $b = \SI{3}{\bit}$, $\rho_\mathrm{AB}\in[0.8,1]$ and $\rho_\mathrm{AE} =\rho_\mathrm{BE} = 0.8$, achieved when Alice, Bob, Eve use uniform quantizers, the GB method, and the ADQC with no quantization error correction transmission, $B = 2$ and $\SI{3}{\bit}$.}
    \label{fig:secretKeyCapacity}
    %\vspace{-.5cm}
\end{figure}
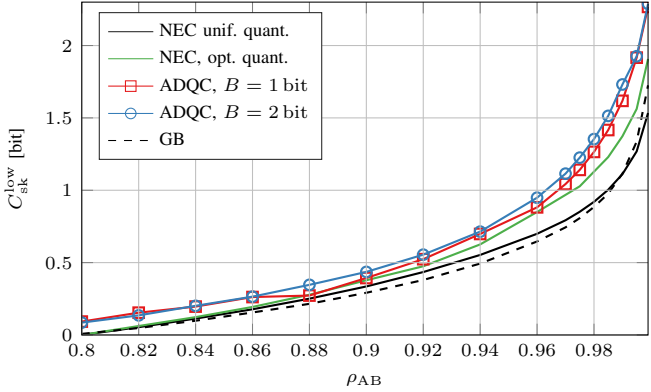

Fig.~\ref{fig:secretKeyCapacity} shows $C_\mathrm{sk}^\mathrm{low}$  for the considered \ac{ska} techniques when extracting $b =3$~bit per sample. We remark that the GB technique discards samples falling on the guard bands, reducing the observation rate (and in general the secret key rate).
%First, we observe that the highest values of $C_\mathrm{sk}^\mathrm{low}$ are achieved when  using \ac{adqc} (i.e., schemes C and E). Thus, sharing information during the advantage distillation is advantageous. 
%ndeed, both the optimization steps \eqref{eq:Eve_optim} and \eqref{eq:AliceBob_optim} improveme the performance of the respective party: 
The best performance is in fact achieved by \ac{adqc} with optimized quantizers, thus, sharing information during the advantage distillation is advantageous. In particular, optimizing the quantizers and using \ac{adqc} yields on average a 60\% improvement of the secrecy capacity, more than doubling it for low correlation values, i.e., when $\rho_\mathrm{AB} \approx \rho_\mathrm{AE} = \rho_\mathrm{BE} = 0.8$. Note that  even the NEC technique with optimized quantizers yields a higher $C_\mathrm{sk}^\mathrm{low}$  with respect to both \cite{graur16} and NEC with uniform quantizers.

\begin{table}
\centering
\caption{Lower bound of the Secret-key capacity achieved with the ADQC, for $\rho_\mathrm{AB}\in[0.8,1]$, $\rho_\mathrm{AE} =\rho_\mathrm{BE}= 0.8$,  $B=\SI{2}{\bit}$, and $b =$2, 3, and $\SI{4}{\bit}$.}
\label{tab:bitPerMeas}
\resizebox{\columnwidth}{!}
{
\begin{tabular}{@{}ccccccccccc@{}}
\toprule
$b\,$[bit] & \multicolumn{10}{c}{$C_\mathrm{sk}^\mathrm{low}\;$[bit]} \\ 
& \cellcolor[HTML]{EFEFEF}0.80 & 0.84 & \cellcolor[HTML]{EFEFEF}0.88 & 0.90 & \cellcolor[HTML]{EFEFEF}0.92 & 0.94 & \cellcolor[HTML]{EFEFEF}0.96 & 0.98 & \cellcolor[HTML]{EFEFEF}0.99 & 0.995 \\\midrule
2 & \cellcolor[HTML]{EFEFEF}0.084 & 0.185 & \cellcolor[HTML]{EFEFEF}0.297 & 0.377 & \cellcolor[HTML]{EFEFEF}0.486 & 0.601 & \cellcolor[HTML]{EFEFEF}0.764 & 1.010 & \cellcolor[HTML]{EFEFEF}1.199 & 1.305 \\
3 & \cellcolor[HTML]{EFEFEF}0.086 & 0.202 & \cellcolor[HTML]{EFEFEF}0.347 & 0.436 & \cellcolor[HTML]{EFEFEF}0.555 & 0.714 & \cellcolor[HTML]{EFEFEF}0.949 & 1.354 & \cellcolor[HTML]{EFEFEF}1.731 & 1.896 \\
4 & \cellcolor[HTML]{EFEFEF}0.095 & 0.247 & \cellcolor[HTML]{EFEFEF}0.314 & 0.414 & \cellcolor[HTML]{EFEFEF}0.577 & 0.779 & \cellcolor[HTML]{EFEFEF}1.039 & 1.455 & \cellcolor[HTML]{EFEFEF}1.867 & 2.305 \\ \bottomrule
\end{tabular}}
\end{table}

Table \ref{tab:bitPerMeas} shows the performance of the \ac{adqc} with $B=\SI{2}{\bit}$ used for quantization error correction and for several values of extracted bit per measurement, $b=2$, $3$, and $\SI{4}{bit}$. Increasing the number of bits extracted from the channel yields a higher $C_\mathrm{sk}^\mathrm{low}$, even just sharing just $B =\SI{2}{\bit}$ of error correction.

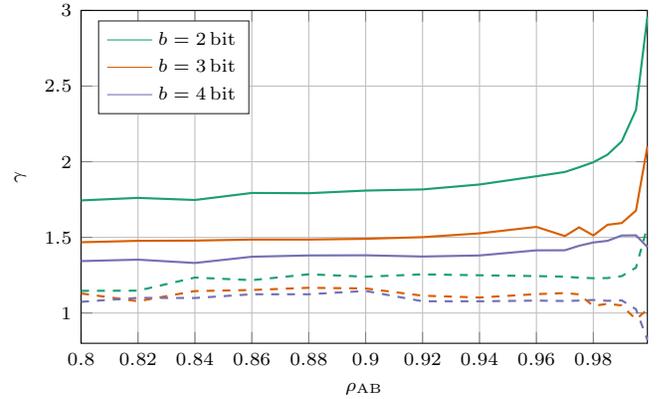
\begin{figure} 
    \centering
    \definecolor{mycolor1}{RGB}{27,158,119}%
\definecolor{mycolor2}{RGB}{217,95,2}%
\definecolor{mycolor3}{RGB}{117,112,179}%

\begin{tikzpicture}[every plot/.style={thick}]

\begin{axis}[%
width=.85\columnwidth,
height=.5\columnwidth,
at={(0\fwidth,0\fheight)},
scale only axis,
xmin=0.8,
xmax=.999,
xlabel style={font=\color{white!15!black}},
xlabel={$\rho_\mathrm{AB}$},
ymin=.8,
ymax=3,
xticklabel style={/pgf/number format/fixed, /pgf/number format/precision=3},
xtick distance = 0.02,
ylabel={$\gamma$},
xmajorgrids,
ymajorgrids,
yminorgrids,
legend style={at={(0.03,0.97)}, anchor=north west, legend cell align=left, align=left, draw=white!15!black},
enlargelimits=false, xlabel style={font=\scriptsize},ylabel style={font=\scriptsize},legend style={font=\scriptsize},ticklabel style={font=\scriptsize}
]
\addplot [color=mycolor1, dashed, forget plot]
  table[row sep=crcr]{%
0.8	1.14653688707357\\
0.82	1.14790006961844\\
0.84	1.23436041873872\\
0.86	1.21722247927163\\
0.88	1.25606437792477\\
0.9	1.24023448349354\\
0.92	1.25550109896235\\
0.94	1.2496299757958\\
0.96	1.24417867908325\\
0.97	1.24020029913786\\
0.975	1.23369737736209\\
0.98	1.23027985639797\\
0.985	1.23119993147042\\
0.99	1.24410277854655\\
0.995	1.30015400100274\\
0.999	1.57045654898724\\
};
\addlegendentry{$b=\SI{2}{\bit}$}

\addplot [color=mycolor2, dashed, forget plot]
  table[row sep=crcr]{%
0.8	1.12948762692449\\
0.82	1.07782550952506\\
0.84	1.14507184628763\\
0.86	1.15226695731717\\
0.88	1.16688155756282\\
0.9	1.1623401820813\\
0.92	1.11461040936912\\
0.94	1.10265342370967\\
0.96	1.12484315181695\\
0.97	1.13202347939703\\
0.975	1.12334775900955\\
0.98	1.04690202557403\\
0.985	1.06125207548198\\
0.99	1.05004938757538\\
0.995	0.958177228261139\\
0.999	1.02352692657022\\
};
\addlegendentry{$b=\SI{3}{\bit}$}

\addplot [color=mycolor3, dashed, forget plot]
  table[row sep=crcr]{%
0.8	1.07401537674148\\
0.82	1.10020338029745\\
0.84	1.09934970065833\\
0.86	1.12397792958541\\
0.88	1.12411644118562\\
0.9	1.14593040444416\\
0.92	1.07779742728708\\
0.94	1.07766545105852\\
0.96	1.082136667627\\
0.97	1.07953594144461\\
0.975	1.0821532393965\\
0.98	1.08631439187009\\
0.985	1.08089349616305\\
0.99	1.08363190328582\\
0.995	1.02499658988756\\
0.999	0.822727713335734\\
};
\addlegendentry{$b=\SI{4}{\bit}$}

\addplot [color=mycolor1]
  table[row sep=crcr]{%
0.8	1.74441323291739\\
0.82	1.76114196567279\\
0.84	1.74720210344429\\
0.86	1.79370971625383\\
0.88	1.7920047357917\\
0.9	1.80941274223863\\
0.92	1.8169728311286\\
0.94	1.84953445935283\\
0.96	1.90392437962172\\
0.97	1.93217138608956\\
0.975	1.96344362153332\\
0.98	1.99654375900643\\
0.985	2.04700549409575\\
0.99	2.13520615063537\\
0.995	2.34294070975604\\
0.999	2.95840553254981\\
};
\addplot [color=mycolor2]
  table[row sep=crcr]{%
0.8	1.46786843852658\\
0.82	1.47715182920221\\
0.84	1.47871953813251\\
0.86	1.48558721354741\\
0.88	1.48558228211679\\
0.9	1.49058729282266\\
0.92	1.50149083166495\\
0.94	1.52658059597832\\
0.96	1.56933173989234\\
0.97	1.5087111649859\\
0.975	1.5663409149326\\
0.98	1.51257583648789\\
0.985	1.58329623320307\\
0.99	1.59481498345633\\
0.995	1.67717702040162\\
0.999	2.09982738760632\\
};
\addplot [color=mycolor3]
  table[row sep=crcr]{%
0.8	1.34325142808243\\
0.82	1.35282450079039\\
0.84	1.33080581109241\\
0.86	1.37210646157231\\
0.88	1.38062382208179\\
0.9	1.38165431059888\\
0.92	1.37343210400687\\
0.94	1.38052986311943\\
0.96	1.41434494741887\\
0.97	1.41475448874784\\
0.975	1.44458708594845\\
0.98	1.4662451275724\\
0.985	1.47722341989681\\
0.99	1.51246991273691\\
0.995	1.51354645782306\\
0.999	1.4364685550609\\
};

%\addlegendimage{empty legend}
%\addlegendentry{\hspace{-.6cm}{\textbf{Scenario:}}}
%\addlegendimage{color = gray}
%\addlegendentry{E}
%\addlegendimage{color = gray, dashed}
%\addlegendentry{D}

%\addlegendimage{empty legend}
%\addlegendentry{\hspace{-.6cm}{$\bm{b=}$}}
%\addlegendimage{color = mycolor1}
%\addlegendentry{\SI{2}{\bit}}
%\addlegendimage{color = mycolor2}
%\addlegendentry{\SI{3}{\bit}}
%\addlegendimage{color = mycolor3}
%\addlegendentry{\SI{4}{\bit}}

\end{axis}
\end{tikzpicture}%
    \caption{Cost $\gamma$ vs correlation $\rho_\mathrm{AB}$ with $\rho_\mathrm{AE} =\rho_\mathrm{BE} = 0.8$, for scenarios $B = \SI{1}{\bit}$ (dashed lines) and $B = \SI{2}{\bit}$ (solid lines), with $b =2$, $3$ and \SI{4}{\bit}. }
    \label{fig:comparision-quant}
    %\vspace{-.5cm}
\end{figure}

We now consider the case of limited side-channel capacity, described in Section~\ref{sec:reconcile}, focusing on the NEC and \ac{adqc} techniques, both with optimized quantizers, to understand the overhead introduced on the side channel. Fig.~\ref{fig:comparision-quant} shows $\gamma$ as a function of the correlation $\rho_\mathrm{AB}$, with $B = \SI{1}{\bit}$,  $b=2$ or $3$, and $B=1$ or $\SI{2}{\bit}$. We first note that for $B=1$ (thus a very limited side-channel overhead due to quantization error correction) the number of bits exchanged on the side channel is very close for both ADQC and NEC schemes (i.e., $\gamma \approx 1$). Indeed, for high values of $\rho_{\rm AB}$ the ADQC technique requires even fewer bits than NEC (for $b=3$ and $4$) since the extracted bit sequences are more similar and the information reconciliation part is less demanding. Instead, when we consider $B=2$, we note that the data rate of the side channel increases by a factor of 3 (for highly correlated channels) to obtain however a higher secrecy capacity as from Fig.~\ref{fig:secretKeyCapacity}.

\section{Conclusion}\label{sec:conclusions}
We have proposed an advantage distillation technique for physical layer-based \ac{ska}, where Alice transmits via a publicly authenticated channel a correction, that is exploited by Bob and, eventually by Eve, to correct their measurements. 
Numerical results show that both the quantizer optimization and the correction transmission allow Alice and Bob to achieve a higher lower bound of the secret key capacity, even when Eve optimizes her quantizers as well. Additionally, we showed that the lower bound of the secrecy key rate per bit shared on the public channel is higher when correction is used, revealing an efficient use of the public channel by this technique.

\bibliographystyle{IEEEtran}
\bibliography{IEEEabrv,biblio}    

\end{document}